# Graphene assisted III-V epitaxy towards substrate recycling

Naomie Messudom(1), Antonella Cavanna(1), Ali Madouri(1), Carlos Macias(2), Nathalie Bardou (1), Laurent Travers(1), Stéphane Collin (1,2), Jean-Christophe Harmand(1), Amaury Delamarre(1,2)

(1) Centre de Nanosciences et Nanotechnologies (C2N), 91120 Palaiseau, France
(2) Institut PhotoVoltaique de France (IPVF), 91120 Palaiseau, France

## Abstract

Re-using the substrate is identified as a method for reducing the cost of high efficiency III-V solar cells. The approach investigated here consists in inserting a graphene layer onto a (001)GaAs substrate prior to the epitaxial growth of GaAs. To obtain a monocrystalline GaAs grown layer, the graphene layer is patterned, followed by a two-step epitaxial growth, here performed by molecular beam epitaxy (MBE). The first step is a selective area growth of GaAs in graphene openings, followed by a lateral overgrowth, under a modulated Ga flux. The second step, after reaching coalescence, consists in a regular growth under continuous Ga supply. It is observed that the pattern orientations relative to the crystallographic direction of the GaAs substrate below the graphene have an influence on GaAs morphology and quality. The best result was obtained for patterns oriented along $[\bar{1}10]+22,5°$ with a graphene coverage of 50%, with a significantly reduced roughness down to 3,3 nm.

## 1. Introduction

To date, III-V solar cells are the most efficient technology available in the market. However, their high cost is a significant barrier to their widespread development. In recent years, a number of studies have been carried out to develop techniques for reducing production costs, such as increasing the growth rate using metal–organic chemical vapour deposition (MOCVD) method[1]. Nevertheless, the most efficient method to reduce the cost of these solar cells is to re-use the substrate which represent the most expensive part in terms of production cost[2]. One of the earliest approaches adopted was chemical epitaxial lift-off, where an AlGaAs or an InGaP sacrificial layer is deposited at the beginning of growth and chemically etched to detach the grown stack. Subsequent techniques such as cleavage of lateral epilayers or spalling techniques[3] have also emerged. Despite the exfoliation process being significantly faster than chemical epitaxial lift-off, there is still a challenge in controlling crack propagation, and the substrate surface remains very rough, requiring additional treatment such as chemical-mechanical polishing (CMP), or the growth of planarization layers, to restore a suitable surface for re-growth. A novel technique known as 2D material-based layer transfer has been the center of much attention[4]. Successful fabrication of single-crystal layers of polar materials on a substrate covered with a monolayer of graphene was demonstrated. It was suggested that a remote interaction, through the graphene, between the underlying substrate and the growing material, allowed obtaining a single crystal aligned with the substrate. However, another alternative mechanism based on defect seeded holes in the graphene followed by lateral overgrowth could also be involved. It was demonstrated to produce single crystalline on both polar and non-polar materials[5], [6]. In this approach, epilayers are seeded in the graphene apertures followed by lateral epitaxial growth on 2D materials[6], [7], [8], [9], [10]. Our previous study reveals that the remote interaction is not observed[11].

To obtain graphene on GaAs, there are several approaches: direct growth of graphene on GaAs or graphene transfer. For direct growth of graphene, the temperature needed are usually too high to preserve the quality of III-

V substrate resulting a poor-quality graphene layer in comparison to others substrate. To transfer graphene on GaAs, dry transfer from Ge or SiC or wet transfer from Cu can be employed. It has been shown that using wet graphene transfer traps oxide at graphene/substrate interface resulting hiding remote epitaxial growth while using a dry transfer process could provide an oxide free surface[12], [13]. The III−V crystal growth for solar cell application needs to smooth and defect free. From now, mainly studies have been done using MOCVD that provides a high growth rates and a high diffusion of III-adatoms[9]. However it has been shown that using a MBE it is possible to growth high quality planar GaAs films on a mask[14].

In this paper, we proposed a growth of GaAs layer on patterned graphene covered (001)GaAs substrate using a two-step growth mode in MBE. A dry transfer method of graphene grown on germanium was previously developed[13]. The investigation focuses first on optimization of growth conditions and pattern geometries to avoid parasitic nucleation at the beginning of the growth process on the graphene, which can affect the quality of GaAs layer. Then, on the influence of patterns orientation on GaAs coalescence to determine the optimal direction for achieving a smooth and defect-free surface with. We show that surface roughness can be reduced to 3,3 nm.

## 2. Experimental details

**Graphene growth and transfer:** The growth of graphene is done on undoped epi-ready Ge(110) by chemical vapor deposition (CVD) method. The substrate is first annealed under H2 at 900°C for 15min then the graphene is grown with a CH4:H2 ratio of 3: 200 sccm. The grown graphene sample is transferred on GaAs substrate using a Ni-assisted dry transfer method. 100nm of Ni is deposited on graphene sample by evaporation. Using a thermal release tape (TRT), we exfoliate both graphene and nickel from Ge and transfer it on GaAs substrate using an air-cushion compression machine. The Nickel is etched in a commercial Transene TFG solution. The remaining Ni-containing residues is cleaned using HCL (37%). This procedure is described in ref [13].

**Graphene patterning**: E-beam lithography is used to pattern the graphene on GaAs substrate. The mask designed to create openings in the graphene layer consists of sets of rectangles with 30 μm as length. There are two sets of rectangles: one with(W) of 150 nm and a period (P) of 300 nm, and the other with the same width but a period of 1 μm. The corresponding graphene coverage for each set is 50% and 85% respectively. For each batch, the patterns have been oriented in 7 directions: [110],[110]+/- 22.5°,[010], [$\bar{1}$10] and [$\bar{1}$10]+/-22,5°.

**Epitaxial growth**: Prior to the growth process, the Gr/GaAs substrate is annealed at 650°C under As pressure for 30 min to remove oxide from the surface. The growth is initiated with periodic supply epitaxy process that will allow a selective area growth of GaAs in graphene apertures. The temperature is measured using an IR Pyrometer with an accuracy of +/-10°. The growth temperature during this step is 610°C with an As/Ga ratio equal to 2. In each cycle, 4ML of Ga is deposited at a growth rate of 0,1ML/s, with a modulated waiting time between Ga pulses. The growth is then followed by a co-evaporation at 580°C, As/Ga ratio equal to 5 and the same growth rate.

**Structural and morphological characterization of the material**: Raman spectroscopy was performed to verify the success of the growth and evaluate the quality of the graphene layer using a Renishaw InVia confocal μ-Raman setup with a 532 nm laser, a 1800 l/mm diffraction grating and a 100x objective operated at 0.72 mW. To obtain topography images of samples, we used an atomic force microscope (AFM) in tapping mode with a silicon probe tip. Scanning electron microscope (SEM) is used for morphological characterization of GaAs before and after coalescence.

## 3. Results and discussion

**Selective area growth of GaAs on patterned graphene using PSE**: To achieve single crystal growth of GaAs on patterned graphene, the process is initiated with a selective area growth of GaAs only on graphene openings. Selective nucleation requires fast diffusion and high desorption of Ga on the graphene layer. The combination of high growth temperature and periodic supply epitaxy (PSE) process has been shown to enhance selectivity[10],

[15], [16]. The growth using PSE protocol consists of sending a periodic pulse of Ga under a constant low As flux as shown in fig. 1a. Interruption time between Ga pulses improves the decomposition of polycrystals on graphene[15]. Fig. 1b shows SEM images of two samples with two different waiting time between Ga pulses: 1 and 2 min. On each sample, we have patterns with the same width (w=150 nm) but two different periods (p=1um and 300nm). For patterns with 1um period, 2 min as waiting time is not enough to achieve a full selectivity of GaAs because there is some parasitic nucleation between slits. Moreover, by decreasing the time between Ga pulses from 2 min to 1 min, the parasitic nucleation between slits and the formation of Ga droplets (inner corner image fig. 1b) increased drastically. However, if the patterns are close, full selectivity is obtained even with 1 min as waiting time. For the following experiments, the period of the patterns is fixed to 300nm and a waiting time of 2 min is applied to ensure a complete selectivity at the first stage of the growth while reducing the growth time.

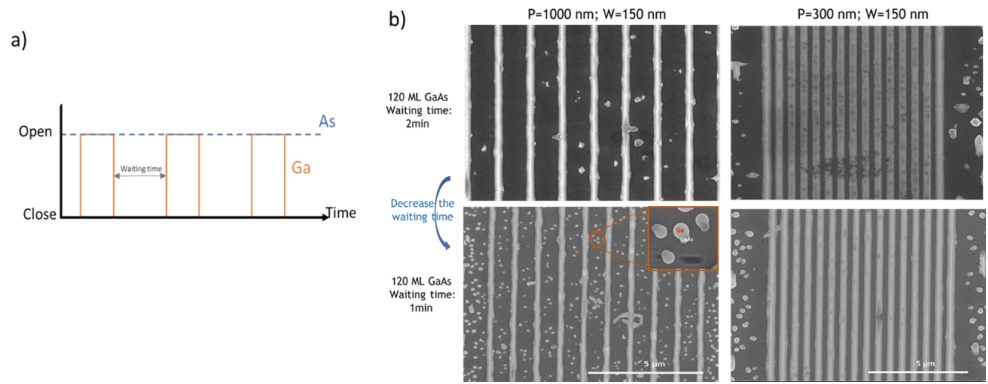

Figure 1: a) PSE process in MBE. c) SAG of GaAs on patterned graphene using PSE process with two waiting times: the first line corresponds to the sample with 2 min as waiting time and the second line, sample with 1 min as waiting time

**Influence of the pattern orientations on GaAs morphology:** To determine the influence of the pattern orientations on GaAs quality, patterns aligned in eight directions have been designed on mask: [110], [110]+22.5°, [110]-22.5°, [010], [$\bar{1}$10], [$\bar{1}$10]+22,5° and [$\bar{1}$10]-22.5°. In crystal structure of GaAs, the planes perpendicular to the surface and intersecting it [110] and [$\bar{1}$10] are symmetric planes, so that [110]+22.5° and [110]-22.5° are equivalent directions, but also [$\bar{1}$10]+22,5° and [$\bar{1}$10]-22,5°. For the results, only one of them is presented. Depending on the orientation of the patterns relative to the crystal direction, different morphology is observed on GaAs. Fig. 2a-e presents SEM images of 120 ML of GaAs growth on patterned graphene with a graphene coverage of 50%, obtained using the PSE process with a waiting time of 2min to ensure a complete selectivity between slits. For apertures along [110] direction, the lateral overgrowth is quite significant with a relatively smooth facet (fig. 2a). By rotating the openings to 22,5° with respect to [110] direction, the facets become rough and the lateral overgrowth decreases (fig. 2d). In [010] direction, the grown GaAs produces the highest lateral overgrowth (fig. 2f) and smooth facets as illustrated in fig. 2c. However, when the apertures are aligned are along [$\bar{1}$10] direction (fig. 2d), the overgrowth is negligible, highlight by the fact that the openings are not complete full and the lateral overgrowth is the smallest (fig. 2f). This behavior can hinder coalescence and merging of GaAs. By rotating the openings by 22,5° relative to [$\bar{1}$10]direction, the lateral overgrowth is observed to increase and exhibits a zig-zag morphology.

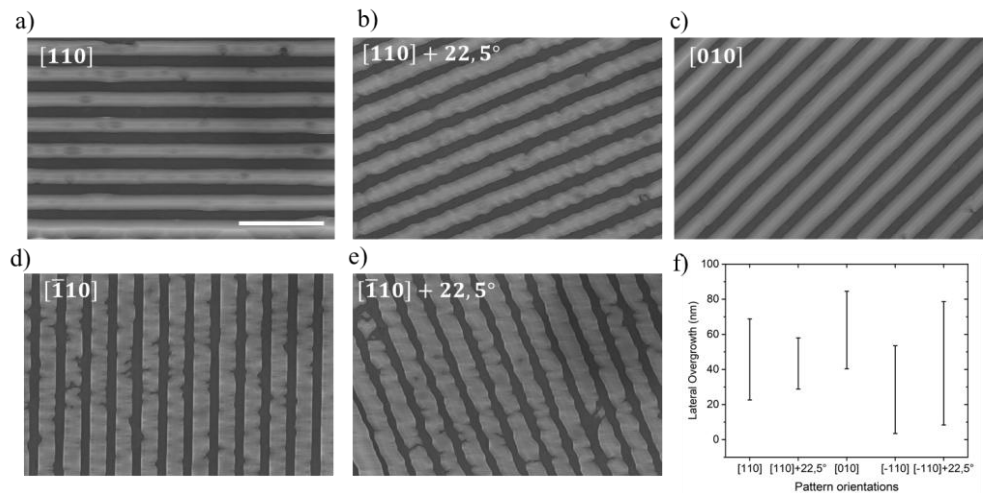

Figure 2: a-e) Results of SAG of GaAs on patterned graphene in 5 directions: [110], [110]+22,5°,[010], [1̄10] and [1̄10]+22,5°. f) Lateral overgrowth along each direction. White lines are epitaxial GaAs growth in the graphene openings and in black area the graphene. P=300nm and W=150 nm. Scale bar:1µm

The PSE process is used until GaAs starts to merge. It corresponds to around 360 ML of GaAs. Then, the second stage of the growth can begin. It is a co-evaporation process where Ga and As flux are sent during the growth. This step allows to achieve coalescence and planarization of GaAs. The thickness of GaAs is 430nm and at the end of the growth the sample is analyzed by SEM and the results are shown in fig. 3. For the growth along the directions of symmetry [110] and [1̄10], GaAs is a very rough and there is quite no coalescence along [1̄10] direction. In the [110] +22,5° direction, the coalescence is better, although a high pit density is observed at the surface. Surprisingly, GaAs growth along the [010] direction which gives the highest lateral overgrowth and the smoothness facets before coalescence results in a rough GaAs surface. Some pits are observed at the surface and it seems the roughness is due to the holes between slits that are not being fully filled. The [1̄10]+22,5° orientation exhibits the best GaAs layer with low defects density and the smoothness surface with a RMS roughness of 3,3 nm (fig. 3f).The result obtained fits well with the typical roughness observed when growing a III-V material on a mask[9], [14], [17].

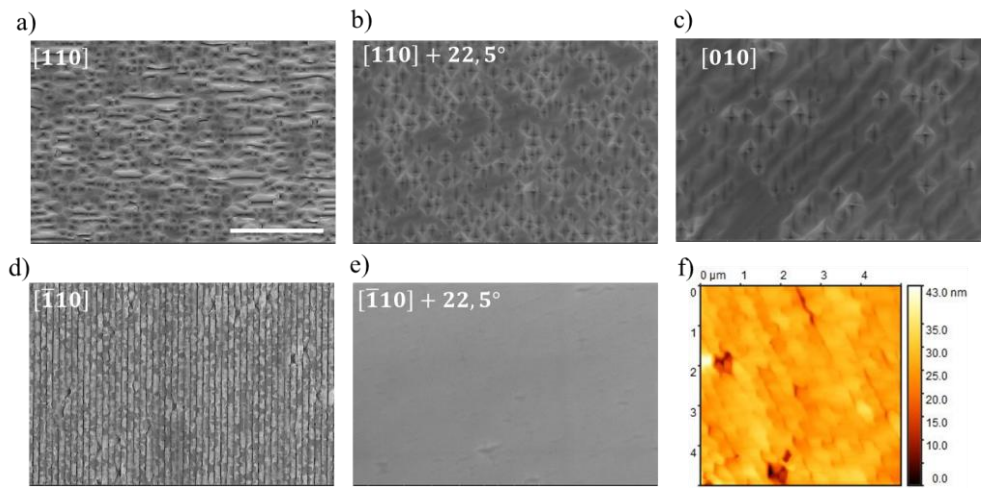

Figure 3: a)-e) SEM images of 500 nm GaAs growth on patterned graphene where patterns are oriented in 5 directions: [110], [110]+22,5°,[010], [$\bar{1}$10] and [$\bar{1}$10]+22,5°. f) AFM image of GaAs in [$\bar{1}$10]+22,5° direction

## 4. Conclusion

We developed an approach to achieve an epitaxial growth of high quality GaAs layer on patterned graphene covered (001)GaAs substrates, with the objective of obtaining an exfoliable epi-layers for substrate re-use in photovoltaic applications. The graphene layer is transferred on GaAs using a Ni-assisted dry transfer method. By creating openings in the graphene layer, it has been able to control nucleation by using a PSE process. It allows to have a selective area growth of GaAs in graphene openings and avoid parasitic nucleation between slits. The size of the patterns and their orientation plays a crucial role in obtaining high quality and planar GaAs epi-layer. We found that there is an orientation dependence of patterns on GaAs morphology and quality. Patterns oriented along [$\bar{1}$10]+22,5° give the best GaAs layer with an average roughness of 3,3 nm closed to epi-ready growth.

## Acknowledgements

This project has been supported by the French Government in the frame of the program of investment for the future (Programme d'Investissement d'Avenir - ANR-IEED-002-01). The authors acknowledge the French Renatech network.